\documentclass[aps,pre,twocolumn,showpacs,showkeys,groupedaddress,floatfix]{revtex4-1}
\usepackage[T2A]{fontenc}
\usepackage[cp1251]{inputenc}
\usepackage[english]{babel}
\usepackage{graphicx}
\usepackage{subfigure}
\usepackage{multirow}
\usepackage{amsmath}
\usepackage{amssymb}
\begin{document}
\title{Specific heat of apple at different moisture contents and temperatures}

\author{Viacheslav Mykhailyk}
\email{mhlk45@gmail.com}
\affiliation{Institute of Engineering Thermal Physics, National Academy of Sciences of Ukraine,
2a, str. Zheljabova, Kyiv 03057, Ukraine
}

\author{Nikolai I. Lebovka}
\email[Corresponding author: ]{lebovka@gmail.com}
\affiliation{Institute of
Biocolloidal Chemistry named after F.D. Ovcharenko, National Academy of Sciences of Ukraine,
42, Boulevard Vernadskogo, 03142 Kiev, Ukraine}

\date{\today}

\begin{abstract}
This work discusses results of experimental investigations of the specific heat, $C$, of apple in a wide interval of moisture contents ($W=0-0.9$) and temperatures ($T = 283-363$ K). The obtained data reveal the important role of the bound water in determination of $C(W,T)$ behaviour. The additive model for description of $C(W)$ dependence in the moisture range of $0.1<W<1.0$ was applied, where the apple was considered as a mixture of water and hydrated apple material (water plasticised apple) with specific heat $C_h$. The difference between $C_h$ and specific heat of dry apple,  $\Delta Cb=C_h-C_d$, was proposed as a measure of the excess contribution of bound water to the specific heat. The estimated amounts of bound water $W_b$ were comparable with the monolayer moisture content in apple. The analytical equation was proposed for approximation of $C(W,T)$ dependencies in the studied intervals of moisture content and temperature.
\end{abstract}

%\pacs{64.60.ah, 64.60.De, 68.35.Rh, 61.43.Bn}

\maketitle

\section{\label{sec:introduction}Introduction}

The knowledge of thermo-physical characteristics of fruit and vegetable tissues is very important for control and evaluation of the quality of foods during their storage and processing. The main thermo-physical characteristics are thermal conductivity $K$, thermal diffusivity $D$ and apparent specific heat $C$. These values are interrelated as $C=K/D\rho$ , where $\rho$  is the density of material \cite{Incropera1996}.

Many previous works were devoted to the study of different thermo-physical characteristics of apples and their dependences on the temperature and moisture content \cite{Gane1936,Sweat1974,Lozano1979, Ramaswamy1979,Ramaswamy1981,  Singh1985,Mattea1986, Ginzburg1987, Mattea1990, Bozikova2006, Bozikova2007, Bozikova2009,Huang2009,Lisowa2000, Lisowa2002, Lisowa2003}. Detailed investigations have shown that temperature and moisture content have significant influence on thermal conductivity $K$, thermal diffusivity $D$ \cite{Riedel1949, Sweat1974, Bozikova2006, Bozikova2007, Bozikova2009, Lisowa2003, Lozano1979, Ramaswamy1981}.

The number of publications, where investigations of the apparent specific heat $C$ of apples are presented, is more limited. The temperature dependence of the apparent specific heat $C$ (kJ kg$^{-1} $K$^{-1}$) in the range of $272$ K$<T<363$ K was approximated by linear relation \cite{Ramaswamy1981, Ramaswamy1979}
\begin{equation}
C=a+b(T-273),                                \label{eq:rel1}\\
\end{equation}
where $a=3.36$, $b=0.0075$ for Golden Delicious apples and $a=3.40$, $b=0.0049$ for Granny Smith apples.
In general, the moisture dependence of specific heat C may be evaluated from the data on total solids and fat content using the additive model \cite{Charm1971}
\begin{equation}
C = C_w + (C_w-C_d)(W-1),	                               \label{eq:rel2}\\
\end{equation}
where $W$ is the moisture content (g H$_2$O/g total), $C_w$ and $C_d$ are specific heats of water ($W=1$,
$C_w\thickapprox 4.187$ kJ kg$^{-1}$К$^{-1}$ at $T=273-373$ К) and dry matter ($W=0$), respectively.

This approximation can be true when the water and dry matter are inert to each other and their mixing or separation is not accompanied by the thermal effects. On the basis of such estimations, the following linear relation was obtained for  the moisture dependence of $C$ (kJ kg$^{-1} $K$^{-1}$) at temperatures close to ambient \cite{Singh1985}
\begin{equation}
C=1.414+0.0272W.                                      \label{eq:rel3}\\
\end{equation}

Here, $W$ is the moisture content (g H$_2$O/g total).
The most problematic in this estimation is the calculation of specific heat of dry matter, $C_d$. Note that typically apples contain water (85.56\%), carbohydrates (11.42\%), and unessential quantity of fibres (2.4\%), proteins (0.26\%), fats (0.17\%) and ashes (0.19\%) \cite{Srikiatden2005}.  The temperature dependencies of specific heat $C$ in different constituents of apples may be found in \cite{Chio2003, Lin2009}. Using the data on specific heat of fresh apples \cite{Ginzburg1987} and of apple juice \cite{Raichkov1983}, the dependence of specific heat $C$ (kJ kg$^{-1} $K$^{-1}$) versus temperature $T$ (within 303-363К) and moisture content $W$ (0.3-0.9) was approximated using the Eq. (1). It was assumed that specific heat of completely dried apple $C_d$ linearly increases with temperature, i.e.
\begin{equation}
C_d(T)=-0.3218+0.00367T.                               \label{eq:rel4}\\
\end{equation}

However, approaches, previously applied for estimation of $C$, are indirect and require the precise data on the content of constituents and studies in the limited ranges of moisture content and temperature.

The purpose of this work was to study the dependencies of the specific heat of apples $C$ in the wide interval of moisture content ($W=0-0.9$) and temperature ($T$=283-263 K). Relations between $C$, $W$ and $T$ and role of the bound water in determination of $C(W,T)$ behaviour are discussed. The analytical equation for approximation of $C(W,T)$ dependencies in the studied intervals of moisture content and temperature is also proposed.

\section{\label{sec:MM}Materials and methods}

\subsection{\label{sec:Mat}Materials}
Fresh "Delikates" apples, obtained from the local market (Kyiv, Ukraine), were used throughout this study. The round slices of  6 mm in diameter and thickness 1-2 mm were cut parallel to the apple axis. Then they were sealed into hermetic aluminium pans. Moisture content was determined after calorimetric measurements. For this purpose, the pans were decapsulated and placed in an oven, where they were kept at 378 K until consecutive weightings, made at 1 h intervals, gave less than 0.5\% variation.

\begin{figure} [htbp]%Figure 1
\centering
\includegraphics[width=0.9 \linewidth]{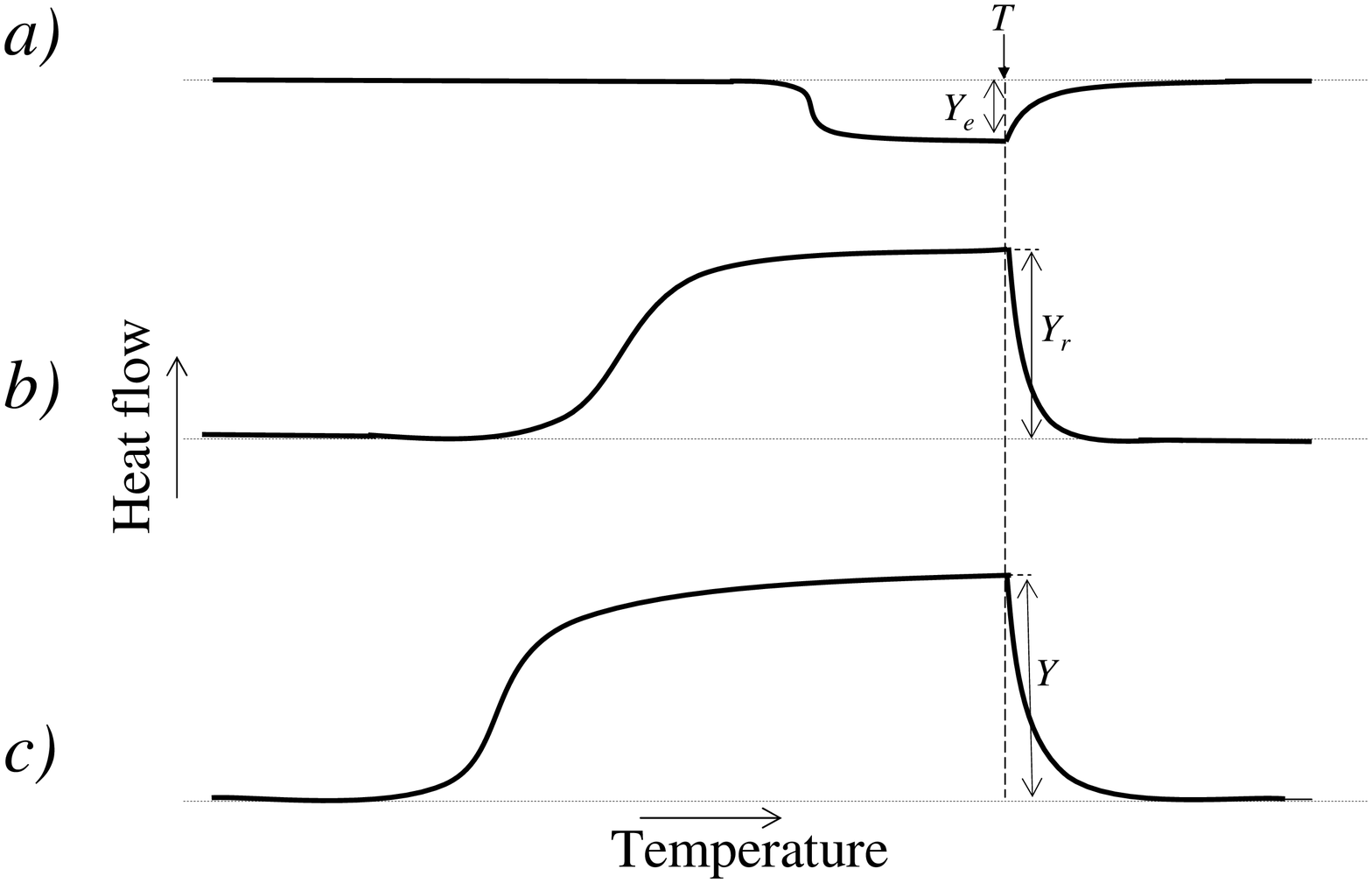}
\caption {\label{fig:comparison}(Color online)
The schematic presentation of a step-by-step protocol, used for specific heat measurements. The deviations of DSC-curves from the basic line during the transition from the scanning regime to the isothermal regime were registered. The measurements were done using the empty container in the comparison cell and the empty container (a), container with the reference specific heat measure (b) and container with investigated sample (c) in the operating cell. Here, $Т$ is the temperature of $C$ value measurement, and $Y$, $Y_r$ and $Y_s$  are the corresponding deviations of DSC curves from the base lines for the experiments a), b and c).}
\end{figure}

\begin{figure*} [!htbp]%Figure 2
\centering
\subfigure[]{\includegraphics[width=0.5\linewidth]{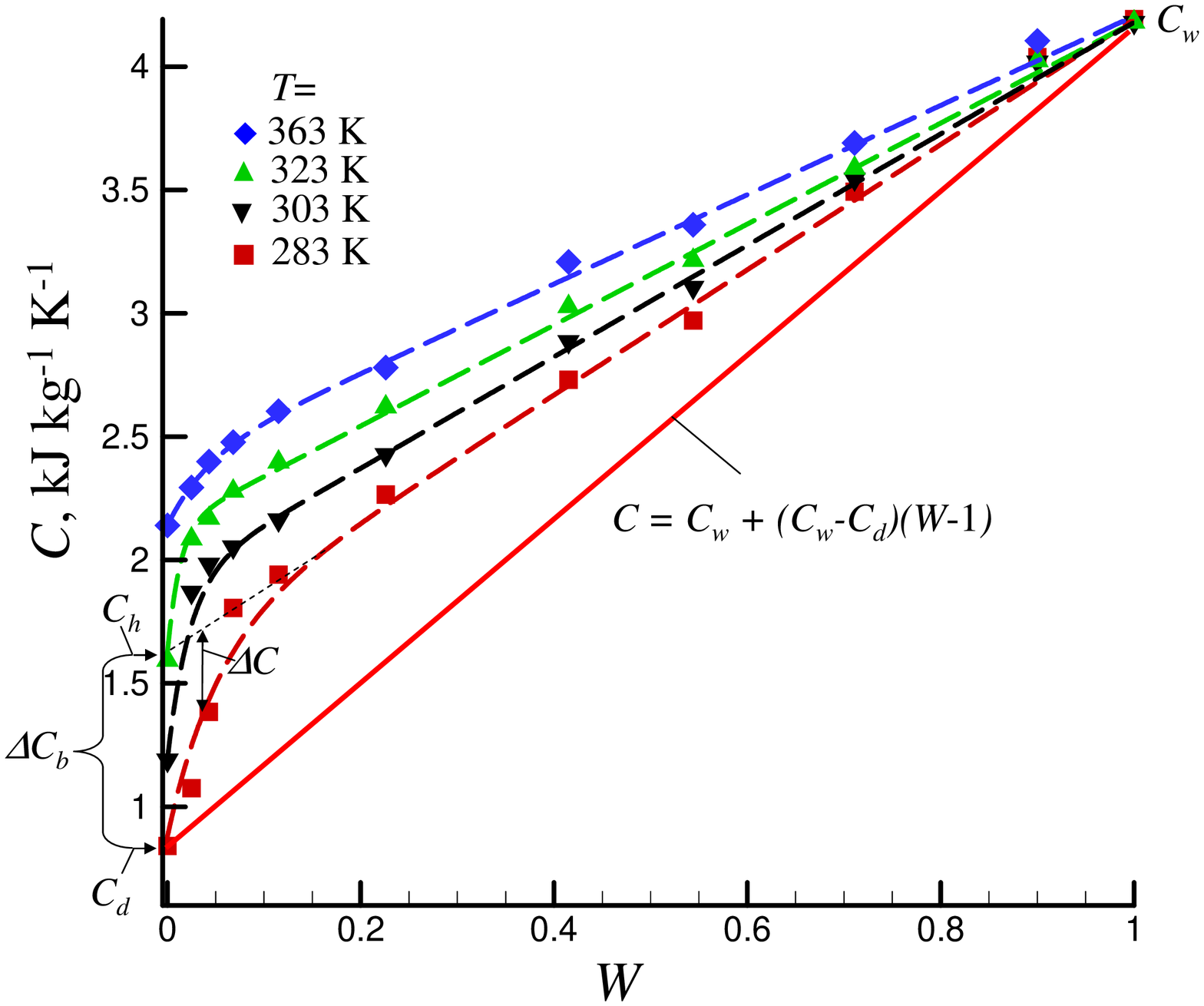}}\hfill
\subfigure[]{\includegraphics[width=0.5\linewidth]{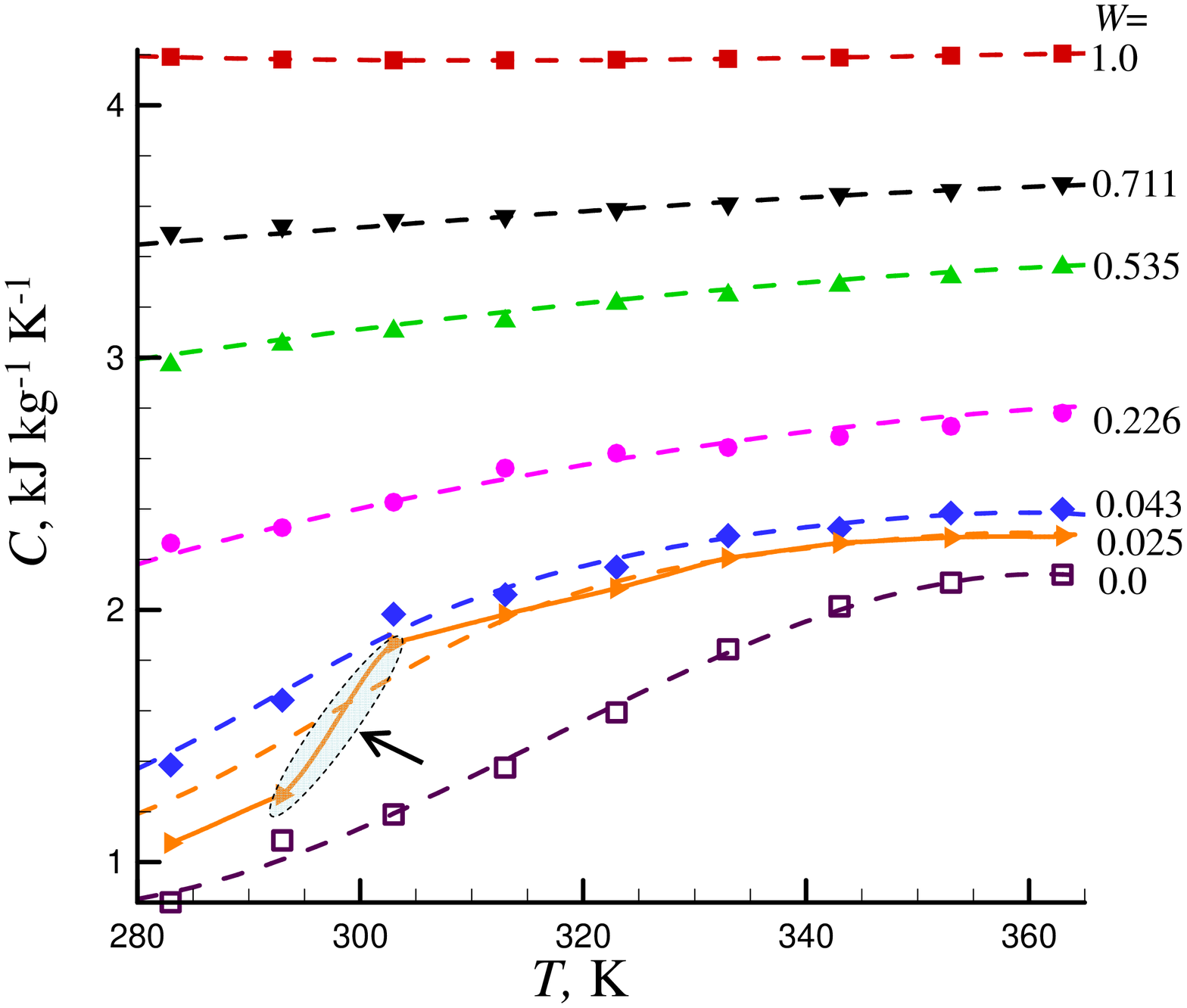}}\hfill
\caption {\label{fig:scaling16}(Color online)
Specific specific heat of apple $C$ versus the moisture content $W$ (a) and temperature $T$ (b).
The example of C(W) data analysis at $T$=283 K is presented in (a). The solid line corresponds to the additive model of specific heat. Here, $C_h$, $C_d$, and $\Delta C_b$ are the contributions of hypothetic hydrated tissue, completely dried tissue, and interactions between water and dry matter of apple, respectively;  $\Delta C$ is the difference between linear contribution $C_l(W)$ and measured value of $C(W)$.}
\end{figure*}

\subsection{\label{sec:model3}Determination of specific heat}
The specific heat was measured using the differential scanning calorimeter DSM-2M, equipped with data logger and software for treatment of data (Specialized Design Office of Instrument Making, Pushchino, Russia). The measuring unit was blown using the drained helium to avoid moisture condensation in the calorimetric cells. The measurements were done using the step-by-step protocols for DSC \cite{Mykhailyk2002}. Such protocol is schematically presented in Fig.1, which includes comparison of deviations of the DSC-curves from the base line during the transition from the scanning regime to the isothermal regime, observed for different samples. The measurements were done using the empty container in the comparison cell and the empty container (a), container with the reference specific heat measure (b) and container with investigated sample (c), in the operating cell.
Samples were heated at 8 K/min in the temperature range of 283 to 363 K. The instrument was calibrated before measurement using synthetic sapphire $\alpha$-Al2O3 as a specific heat standard reference material \cite{Sabbah1999}. The specific heat of the sample $C_s$ was calculated from the equation
\begin{equation}
C_s=C_r[(Y_s-Y)m_r]/[(Y_r-Y)m_s],                               \label{eq:rel5}\\
\end{equation}
where $C_r$ is the specific heat of the reference, $m_s$ and $m_r$ are the masses of the sample and reference, respectively, $Y$, $Y_r$ and $Y_s$ are the corresponding deviations of DSC curves from the base lines for the experiments a), b), and c).

The measurement accuracy was estimated by application of the step-by-step protocols for determination of specific heat of the reference sample (sapphire). The mean square error and relative error were 0.155 J kg$^{-1}$ K$^{-1}$ and 0.67 \%, respectively.

\subsection{\label{sec:stat}Statistical analysis}
Each experiment was repeated, at least, three times. One-way analysis of variance was used for statistical analysis of the data with the help of Statgraphics plus (version 5.1, Statpoint Technologies Inc., Warrenton, VA). Significance level of 5 \% was assumed for each analysis. The error bars, presented on the figures, correspond to the standard deviations.

\section{\label{sec:results}Results and their discussion}

\begin{table*}[!htbp]
  \caption{Specific thermal capacity of apple $C$ (kJ kg$^{-1}$ K$^{-1}$) at different water contents, $W$, and temperatures, $T$.}\label{tab:tab1}
\begin{ruledtabular}
  \begin{tabular}{cccccccccc}
$W$,(g H$_2$O/g total) &  $283 $K  &  $293 $K  &  $303 $K  &  $313 $K &  $323$ K &  $333$K &  $343$ K &  $353$ K &  $363$ K \\
\hline
0.000 &     0.841&	1.086&	1.189&	1.375&	1.594&	1.844&	2.014&	2.108&	2.140\\
0.025 &     1.076&	1.267&	1.869&	1.982&	2.086&	2.206&	2.266&	2.287&	2.294\\
0.043 &     1.385&	1.642&	1.983&	2.060&	2.169&	2.294&	2.322&	2.384&	2.399\\
0.068 &     1.806&	1.958&	2.053&	2.144&	2.279&	2.370&	2.410&	2.441&	2.478\\
0.115 &     1.941&	2.061&	2.163&	2.245&	2.397&	2.494&	2.508&	2.547&	2.603\\
0.226 &     2.265&	2.326&	2.427&	2.562&	2.621&	2.644&	2.687&	2.728&	2.780\\
0.415 &     2.731&	2.816&	2.885&	2.982&	3.027&	3.052&	3.110&	3.169&	3.208\\
0.544 &     2.971&	3.053&	3.105&	3.144&	3.215&	3.247&	3.286&	3.320&	3.359\\
0.711 &     3.493&	3.522&	3.544&	3.560&	3.588&	3.610&	3.647&	3.663&	3.690\\
0.900 &     4.038&  4.038&	4.018&	4.022&	4.024&	4.020&	4.056&	4.071&	4.104\\
1.0 (water)&4.192&	4.182&	4.179&	4.179&	4.181&	4.184&	4.190&	4.196&	4.205\\
  \end{tabular}
  \end{ruledtabular}
\end{table*}

Table 1 presents the measured data of the specific heat of apple $C$ versus water content, $W$, and temperature, $T$. For the sake of clarity, some of these data are presented in the form of $C(W)$ (Fig. 2a) and $C(T)$ (Fig. 2b) dependencies. It is remarkable that $C(W)$ dependencies were practically linear at $W>0.1$ and non-linear at small humidity ($W< 0.1$). The linear contribution to the specific heat, $C_l$, may be described as
\begin{equation}
C_l = C_w + (C_w-C_h(T))(W-1),                                \label{eq:rel6}\\
\end{equation}
where, $C_h$ is the value of $C_l(W)$ intercept (See, insert in Fig. 2a).

The observed behaviour is not surprising accounting for the fact that water in apples has heterogeneous structure and can be divided into intercellular or capillary water (free water), multilayer water (weakly bound water), and monolayer water (tightly bound to the polar sites of apple tissues) \cite{Okos1992}. The studies of moisture sorption isotherms evidence that the water content, associated with the monolayer moisture content of fresh apple, varies from $W$= 0.039 to 0.084 \cite{Mrad2012}. The maximum moisture content, corresponding to the monolayer moisture in fruits, ranges from 0.1 to 0.17 \cite{Lim1995}, which is in accordance with transition between linear and non-linear dependence $C(W)$, observed at $W\approx 0.1$.
It is remarkable also that a characteristic jump in $C(T)$ was observed at certain critical temperature, $T\approx 290$ K, at small water content (including 0.02-0.03 of bound water) (Fig. 2b).  Such behaviour is similar to that of temperature dependence of specific heat in denatured biopolymers and may be identified as glass transition
\cite{Tsereteli1989, Tsereteli1990}. It reflects the transition from a relatively stable glassy state to a metastable rubbery state \cite{Rahman2006}. The glass transition temperature $T_g$ increases with decrease of moisture content and the value of $T_g\approx 290$ K at $W=0.023$ is in correspondence with $T_g(W)$ dependencies, previously reported for apple \cite{Mrad2012}. If water content is small, the apple is in an amorphous metastable state, which is very sensitive to changes in moisture content.

The observed linear behaviour of specific heat in the moisture range of $0.1< W<1$ evidences applicability of the additive model for the mixtures of water and hypothetic hydrated apple material (water plasticised apple) with the specific heat $C_h$. At small moisture content, the systematic deviation between linear contribution $C_l(W)$  and measured value of $C(W)$ was observed. It reached maximum,  $\Delta C= \Delta C_b=C_h-C_d$, in completely dried apple (at $W=0$, see, insert in Fig. 2a).

The measured values of $C$ always noticeably exceeded the prediction of the additive model, Eq.(2), (see, deviations between data point and solid line for $T=283$ K in Fig. 2a). It evidences the supplementary contribution of the, so called, bound water to the specific heat of hydrated apple. The value of  $C_b$ characterises the excess contribution of interactions between water and dry matter and the effect of water confinement in pores of apples. The concept of bound water is widely accepted and accounts for the possibility of structural changes in the regular structure of water in hydration shells \cite{Galamba2013, Li2007} and anomalous behaviour of the specific heat of water confined to pores \cite{Nagoe2010}.

\begin{figure} [htbp]%Figure 3
\centering
\includegraphics[width=0.9 \linewidth]{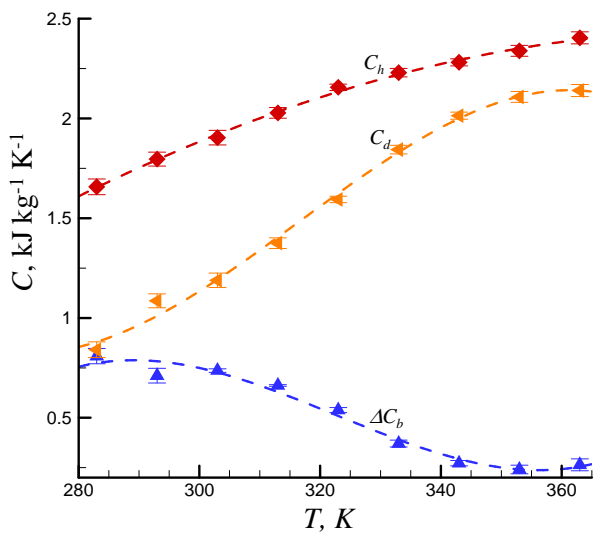}
\caption {\label{fig:f03}(Color online)
The temperature dependences of different specific heat components $C_h$, $C_d$ and $\Delta C_b$, related to contributions of hypothetic hydrated tissue, completely dried tissue and bound water, respectively.
}
\end{figure}

The temperature dependences of the specific heat components $C_h$, $C_d$ and  $C_b$ are shown in Fig. 3. Note that contributions $C_h$, $C_d$ grow and the excess contribution $\Delta C_b$ decreases as temperature $T$ increases. The observed decrease of  $\Delta C_b$ with temperature evidences diminution of the excess contribution of bound water and may be explained by the damage of the specific structure of bound water. The excess values of $\Delta C_b$ were rather small (0.25-0.8 kJ kg$^{-1}$ K$^{-1}$) in comparison to that of free H$_2$O ($C_w =4.187$ kJ kg$^{-1}$ K$^{-1}$). However, the contribution of bound water was crucial for violation of the additive model, Eqs. (2,6).

Usually, the additive model (Eq. 2) is applied for estimation of the effective specific heat of bound water $C_b^e$
\cite{Bull1968, Suurkuusk1974, White1960}. The additive approximation for evaluation of $C_b^e$ may be rewritten as
\begin{equation}
C/(1-W)=C_d +C_b^e(W/(1-W)).                               \label{eq:rel7}\\
\end{equation}

The plot of experimental data presented as $C/(1-W)$ against $W/(1-W)$ gives the $C_b^e$ value as a slope.

\begin{table*}[!htbp]
  \caption{Coefficients of polynomial approximations of the temperature dependencies of $C_w(T)$,$C_h(T)$, $C_b(T)$,$W_b(T)$ and $C_d(T)$ (Eq. 8) with respective coefficients of determination$\rho$.
}\label{tab:tab2}
\begin{ruledtabular}
  \begin{tabular}{cccccccccc}
$ $     &     $C_w$&            $C_h$&              $C_b$&              $W_b$ &             $C_d$             \\\hline
$a$     &     8.5834614350E+00&	-6.2058771600E+00&	-1.1766931000E+02&	3.1565498570E+00&	9.8574175420E+01  \\
$b$     &     -3.7897960000E-02&3.6690250000E-02&   1.1233533270E+00&	-2.2407320000E-02&	-9.6523008000E-01 \\
$c$     &     1.0704300000E-04&	-1.6304000000E-05&	-3.5194100000E-03&	4.8544900000E-05&	3.1232050000E-03  \\
$d$     &     -9.8822000000E-08&	-5.3715000000E-08&	3.6355500000E-06&	-2.8809000000E-08&	-3.2946000000E-06\\
$\rho$  &     0.9937&	0.9976&	0.9773&	0.8607&	0.9950\\
  \end{tabular}
  \end{ruledtabular}
\end{table*}

The contribution of bound water to the specific heat was previously observed in hydrated state of different materials, e.g., poly(acrylic acid) \cite{Xu2011}, proteins \cite{Gasan1994, Yang1979}, various foods \cite{Riedel1966}, food gels \cite{Cornillon1995}, and concretes \cite{Bentz2011, Tatro2006, Waller1996}. The obtained data were rather divergent and evidenced that contribution of the bound water to the specific heat was dependent upon the nature of hydrated materials. E.g., it was demonstrated that the most tightly bound water in lysozyme was characterized by the specific heat of 2.3 kJ kg$^-1$ K$^-1$ , a value close to the specific heat of ice \cite{Yang1979}. For bound water, incorporated into the product of concrete hydration, the estimates gave significantly reduced value of specific heat, approximately 0.91-2.2 kJ kg$^{-1}$ K$^{-1}$ \cite{Bentz2011, Tatro2006, Waller1996}. However, the contribution of bound water in collagen to the specific heat was approximately 5.35 kJ kg$^{-1}$ K$^{-1}$  (i.e., higher than that of free water) over the entire range of water content from 1 to 100\% \cite{Hoeve1976}. According to predictions of computer simulations, specific heat of water hydration shells in peptides exceeds C value in bulk and decreases upon heating
\cite{Oleinikova2010}.

\begin{figure} [htbp]%Figure 4
\centering
\includegraphics[width=0.9 \linewidth]{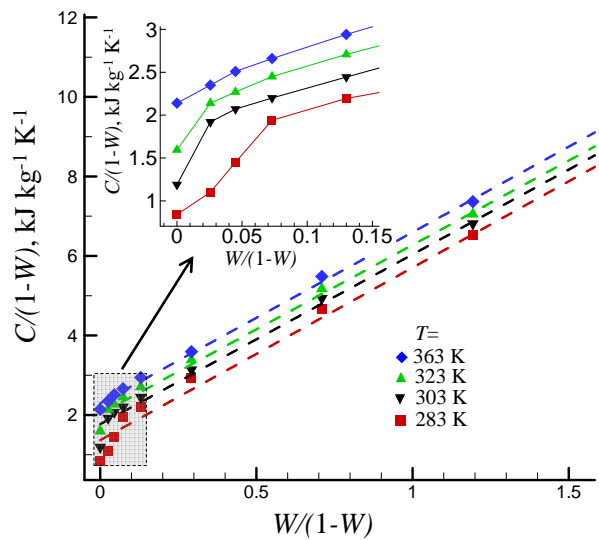}
\caption {\label{fig:f04}(Color online)
To the evaluation of effective specific heat of bound water $C_b^e$ using Eq. (6). The dashed lines are the results of linear fitting over the whole interval of moisture content ($W=0-0.9$). The insert show the $C/(1-W)$ versus $W/(1-W)$  dependencies at small moisture content ($W=0-0.115$).
}
\end{figure}

Figure 4 present the experimental data of this work in coordinates $C/(1-W)$ against $W/(1-W)$. These dependencies were noticeably nonlinear at small moisture content (See insert in Fig. 4). So, it is evident that results of $C_b^e$ estimation can be strongly dependent on the choice of interval of moisture content. In principle, the direct calculation of the effective specific heat of water $C_b^e$ using the additive model can result in significant overestimation or underestimation of the specific heat of bound water.

\begin{figure} [htbp]%Figure 5
\centering
\includegraphics[width=0.9 \linewidth]{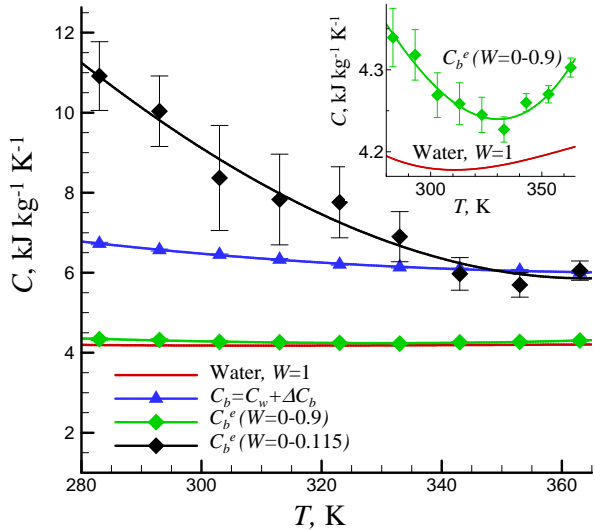}
\caption {\label{fig:f05}(Color online)
Temperature dependencies of the specific heat of pure water and different estimations of $C_b^e$, and  $Cb=C_w+\Delta C_b$. Insert compares the thermal capacity data for pure water ($W=1$) and $C_b^e$($W=0-0.9$).
}
\end{figure}

Figure 5 compares the temperature dependencies of the specific heat of pure water $C_w$ and different estimations of $C_b^e$, and $C_b=C_w+ \Delta C_b$. The estimation of $C_b^e$ using the experimental data within the interval of small moisture content ($W=0-0.115$) results in larger values of specific heat as compared with $C_b=C_w+ \Delta C_b$. However, the difference between $C_b^e$ and $C_b=C_w+ \Delta C_b$ becomes smaller with temperature increase, and both estimations give approximately the same result at $T>340$ K. For wider interval of moisture content ($W=0-0.9$), the effective specific heat of water $C_b^e$ was insignificantly larger than that of pure water $C_w$. However, it is remarkable that temperature dependence of $C_b^e$ demonstrated the presence of a rather deep minimum at $T\approx 330-340$ K. The similar minimum in pure water is observed at $T \approx 310.7$ K.

\begin{figure} [htbp]%Figure 6
\centering
\includegraphics[width=0.9 \linewidth]{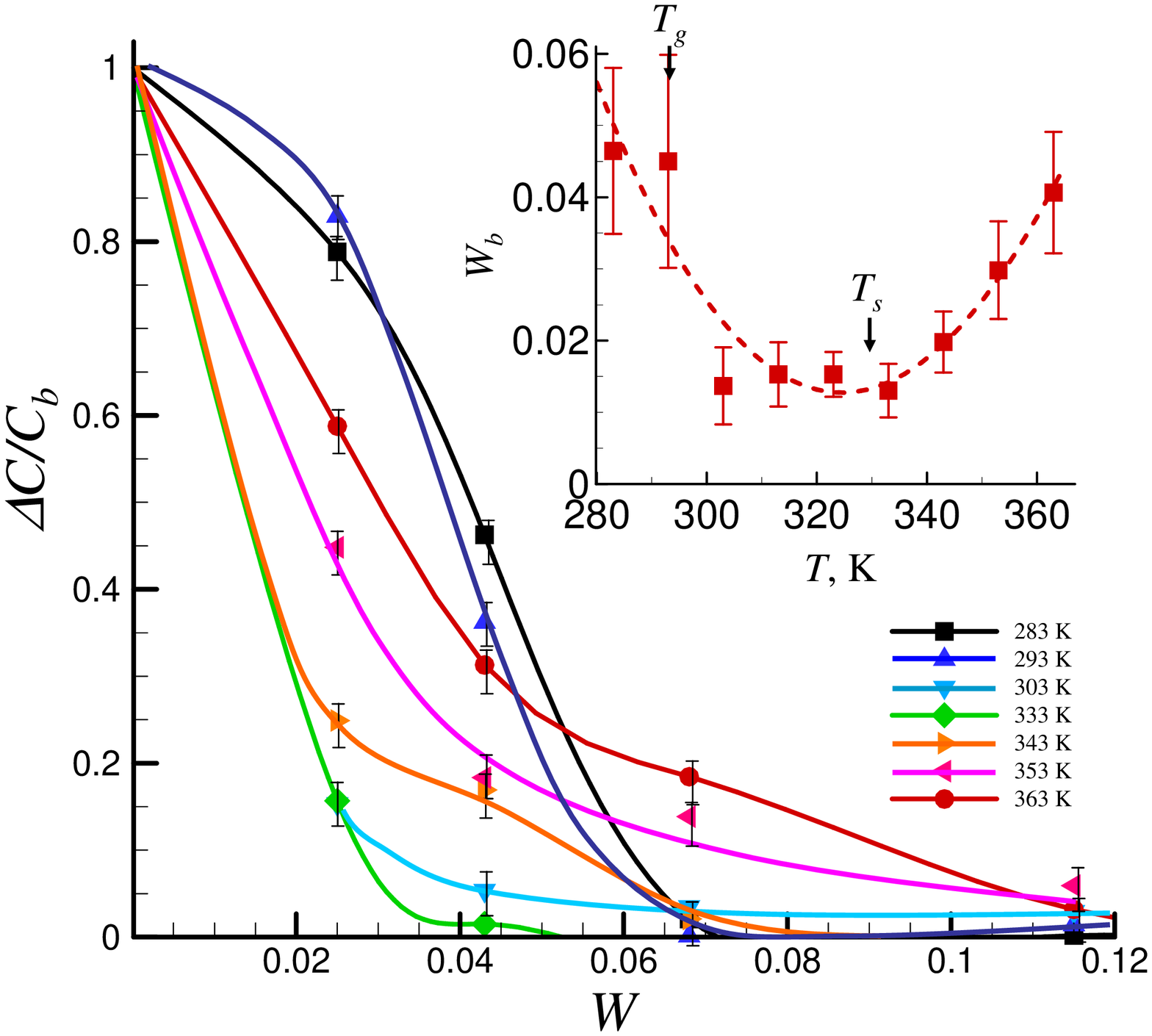}
\caption {\label{fig:f06}(Color online)
The ratio $\Delta C/\Delta C_b$= versus moisture content $W$ at different temperatures, $T$. Insert shows the amount of bound water $W_b$ versus temperature $T$.}
\end{figure}

The amount of bound water $W_b$ can be estimated from analysis of  $\Delta C/\Delta C_b$ versus moisture content $W$ dependencies at different temperatures $T$ (Fig. 6). The value  $\Delta C/\Delta C_b$ rapidly decreases as $W$ increases and approaches zero at $T> 0.1-0.15$. These dependencies can be used for estimation of the amount of bound water $W_b$ in the apple tissue. The value of $W_b$ can be roughly estimated from fitting of $\Delta C/\Delta C_b$ using the exponential function:
\begin{equation}
\Delta C/\Delta C_b=\exp(-W/W_b).                               \label{eq:rel8}\\
\end{equation}

Insert in Fig. 4 shows the temperature dependence of bound water content $W_b$. The estimated values of $W_b$ are comparable with those corresponding to the monolayer moisture content of fresh apple, $W$= 0.039 to 0.084 \cite{Mrad2012}, and pass through the minimum as the temperature increases. The decrease of $W_b$ above $T_g=290$ K, possibly, reflects the presence of glass transition, at which a solid ''glass" transforms into a liquid-like ''rubber" \cite{Rahman2006}. It can be speculated that extend of water structure perturbation by the apple is higher in the glass state. Note that decrease of the monolayer moisture content with temperature increase in the interval of $T=303-333$ K was observed earlier \cite{Mrad2012}.  From the other side, the observed increase of $W_b$ above $T_s\approx 330$ K can reflect the thermally induced softening of apple tissue at elevated temperatures. However, the precise mechanism of such behaviour is still unclear and requires supplementary investigations.

Finally, from the practical point of view, the following approximation for the moisture and temperature dependencies of the specific heat of apple fulfils:
\begin{widetext}
\begin{equation}
C=C_w(T)+(C_w(T)-C_h(T))*(W-1)- C_b(T)*\exp(-W/W_b(T)),                               \label{eq:rel9}\\
\end{equation}
\end{widetext}
where all the temperature dependencies $C_w(T)$,$C_h(T)$, $C_b(T)$,$W_b(T)$ and $C_d(T)$ can be roughly approximated using the polynomials:
\begin{equation}
y(T) = a+bT+cT^2+dT^3,	                               \label{eq:rel10}\\
\end{equation}
where values of $a, b, c, d$ and of the coefficients of determination $\rho$ are presented in Table 2.
The dashed lines in Figures 2, 3 and 6 were obtained using the Eq. 8, Eq. 9 and coefficients presented in Table 2. It can be seen that Eqs. (9)-10 allow satisfactory approximation of the experimentally observed $C(W,T)$ dependencies with exception of $C(T)$ dependence for $W=0.025$ in the vicinity of glass transition temperature,  $T_g\approx 290$ K.

\section{\label{sec:concl}Conclusion}

The detailed study of specific heat at different water contents, $W$, and temperatures, $T$ revealed the important role of bound water in determination of the thermal properties of apple. The observed $C(W)$ dependencies were practically linear at $W> 0.1$ and non-linear at small humidities ($W<0.1$). The nonlinear dependencies reflected the specific heat behaviour at monolayer and sub-monolayer moisture contents, where the effects of bound water were pronounced. The characteristic jump in $C(T)$ behaviour at small water content ( 0.02-0.03 of bound water), probably, reflects glass transition in water plasticised apple material. The additive model can be satisfactory applied for description of C(W) dependence in apple in the moisture range of $0.1<W<1$ considering apple as a mixture of water and hydrated apple material (water plasticised apple) with the specific heat $C_h$. The difference between $C_h$ and specific heat of dry apple,  $\Delta C_b=C_h-C_d$, can be treated as a measure of the excess contribution of bound water to the specific heat. The observed decrease of  $\Delta C_b$ with temperature increase evidences diminution of the excess contribution of bound water and may be explained by the damage of specific structure of bound water. Analysis of the data has shown that commonly applied direct calculations, based on the additive model, can result in significant errors in estimation of the specific heat of bound water. The estimated amounts of bound water $W_b$ were comparable with the monolayer moisture content in apple and were passing through the minimum as the temperature was increasing. Such behaviour can be explained accounting for the glass transition at $T_g\approx290$ K and thermal softening of apple tissue at $T>T_s=330$ K. The proposed Eqs. (9) and (10) can be useful for analytical calculation of $C(W,T)$ in the wide interval of moisture content ($W=0-0.9$)  and temperature ($T=283-363$ K).

\begin{acknowledgments}
The authors appreciate the financial support from the National Academy of Sciences of Ukraine. Authors also thank Dr. N. S. Pivovarova for her help with the preparation of the manuscript.
\end{acknowledgments}

\end{document}